\input psfig.tex
\documentstyle[a4,12pt,epsf]{article}

\begin{document}
\baselineskip .80cm

\vbox{\vspace{6mm}}

\begin{center}{
{\large \bf
INCORPORATION OF SPACETIME SYMMETRIES IN EINSTEIN'S FIELD
EQUATIONS}\\[7mm]
Elias Zafiris\\
{\it Theoretical Physics Group \\
Imperial College \\
The Blackett Laboratory \\
London SW7 2BZ \\
U.K. \\
e.mail:e.zafiris@ic.ac.uk \\}
\vspace{2mm}
IMPERIAL/TP/96-97/06, PACS 04
}\end{center}
\vspace{4mm}
\begin{abstract}
In the search for exact solutions to Einstein's field equations the
main simplification tool is the introduction of spacetime
symmetries. Motivated by this fact we develop a method to write the
field equations for general matter in a form that fully incorporates
the character of the symmetry. The method is being expressed in
a covariant formalism using the framework of a double congruence. The
basic notion on which it is based is that of the geometrisation of
a general symmetry. As a special application of our general method  we
consider  the case of a spacelike conformal Killing vector field on
the spacetime manifold regarding special types of matter fields. New
perspectives in General Relativity are discussed. 
\end{abstract}

\section{Introduction}
In recent years there has been a lot of research work in symmetries in
General Relativity. Originally, the motivation was the need to
simplify Einstein's field equations in the search for exact solutions,
and the introduction of symmetries or collineations served as the
basic tool.

The types of symmetries dealt with are those which arise from the
existence of a Lie algebra of vector fields on the spacetime manifold
which are invariant vector fields of certain geometrical objects on
this manifold. The symmetries can be expressed in relations of the
form $L_{\xi}W=Y$ where $W$ and $Y$ are two geometrical objects on the
spacetime manifold and $\xi^{a}$ is the vector field generating the
symmetry [1].

The most important and common symmetries are those for which $W$ and
$Y$ are one of the fundamental tensor fields of Riemannian geometry,
namely $$g_{ab},\quad {\Gamma^{a}}_{ bc},\quad R_{ab},\quad
R_{abcd}$$
A diagram defining these symmetries and giving their relative
hierarchy is given in [2].

The most extensively studied symmetries in the context of General
Relativity are the Killing vectors and the Conformal Killing vectors
[3,4,5,6,7]. Some work has also been done on Affine collineations [8,9,10],
Curvature collineations [2,11,12] and contracted Ricci collineations
[13,14,15]. The latter symmetries are much more difficult to be
studied, because besides the metric they involve other geometrical
objects which obey concrete conditions, that make the hadling and the
interpretation of the equations difficult. A unified approach to all
these symmetries from a purely differential geometric point of view has been recently given in [16].

It seems to us that although the motivation to study all the above
mentioned symmetries was the simplification of Einstein's equations in
the search for exact solutions, this initial aim has been partially
unfulfilled. More concretely there doesn't seem to appear in the
literature, so far as we know, a general framework which permits the
incorporation of a symmetry of every possible kind in Einstein's
equations for general matter.

Towards this direction we develop a method to write the field
equations for general matter in a form that fully incorporates the
character of the symmetry. Our method is based on the notion of
geometrisation of a general symmetry, that is we employ the
description of it as necessary and sufficient conditions on the
geometry of the integral lines of the vector field which generates the
symmetry. Attempts towards this direction have been done for timelike Conformal Killing vectors [17,18], using the theory of timelike congruences [19,20,21,22,23], for spacelike
Conformal Killing vectors [24], using the theory of spacelike
congruences [25,26], and recently a step towards a generalisation of
this idea has been presented in [27].

The most important aspect of the ``geometrisation'' method to study a
symmetry is that the symmetry is expressed in a form that it is suited
to the simplification of the field equations in a direct and inherent
way.

The method we develop may be outlined as follows:

The introduction of a symmetry is most conveniently studied if we
consider the Lie derivative of the field equations with respect to the
vector $\xi^a$ which generates the symmetry. After doing this we
obtain an expression  which on
the left-hand side contains the Lie derivative of the Ricci tensor and
on the right-hand side contains the Lie derivative of the energy
momentum tensor. Eventually, we obtain the field equations as Lie
derivatives along the symmetry vector of the dynamical variables. The
$L_{\xi}R_{ab}$ can be calculated directly from the symmetry in terms
of $\xi_{a;b}$. If we employ the corresponding theory of congruences
we can express the  $\xi_{a;b}$ in terms of the kinematical quantities
(expansion, vorticity, shear) which characterise the congruence
generated by the symmetry vector. The next step is to use the expression of
a general symmetry as an equivalent set of conditions on the
kimematical quantities characterising the congruence, namely we apply
the geometrisation of a symmetry. If we substitute these conditions
into the general expression for $L_{\xi}R_{ab}$ we manage to write
directly the field equations, for any type of matter, in a way that
they inherit the symmetry of  $\xi_{a}$.

In this work we compute the  $L_{\xi}T_{ab}$ using the most general
form of the energy-momentum tensor $T_{ab}$, and the method is being
demonstrated assuming only that $\xi_{a}$ is a spacelike vector
orthogonal  to the 4-velocity $u^{a}$ of the observers. This choice is
justified by the fact that we wish to keep the length of the equations
as short as possible and at the same time exhibit all the steps of the
method in a transparent manner, without falling into irrelevent to our
purposes complications. Although we work with a spacelike symmetry
vector it is plausible that our approach can be applied to an
arbitrary (non-null) $\xi_{a}$, tilted with respect to $u^{a}$. (We
note that the case in which  $\xi_{a}$ is timelike is much more
simpler).

Having considered general matter it is evident that all the studies
referring to various simplified types of matter like perfect fluids,
charged fluids, anisotropic fluids and so on, consist special cases of
our general scheme and they offer the ground to check the validity of
our results. Thus we finally manage to recover and extend the results
of the current literature, for example those of references [28,29,30] and [31].

The structure of the paper is as follows:

In section 2 we introduce the double congruence covariant framework
permitting the introduction of arbitraty reference frames. In 
section 3 we present the generic formulation of symmetries in General
Relativity and then we study the geometrisation of spacetime
symmetries. The key formal results of the paper are contained in
section 4 where we construct the symmetries-incorporated Einstein's
field equations for general matter. Section 5 shows how these results
are used in the situation of a spacelike Conformal Killing vector
symmetry discussing  the perfect fluid 
case. Finally we summarize and conclude in section 6, discussing
further avenues of research. The applications considered here are
taken just far enough to demonstrate and provide a familiar context
for the techniques developed in this paper. I expect to return in
later papers to the new applications which these techniques make possible.

The notation will be the usual one. {\bf M} will denote the spacetime
manifold with metric {\bf g} of Lorentz signature (--,+,+,+), which is
assumed to be smooth. The Riemann, and Ricci tensors are denoted
by $R_{abcd}$, and $R_{ab}$, respectively, whilst a
semi-colon denotes a covariant derivative, a comma a partial
derivative and L is the Lie derivative. Round and square brackets will
denote the usual symmetrisation and skew-symmetrisation of
indices. Latin indices take the values 0,1,2,3 and units
are used for which the speed of light and Einstein's gravitational
constant are both unity.

\section{The double congruence framework}

In order to develop our method we are going to use the most general
approach to the description of reference frames, namely the theory of
congruences. This approach is explicitly general covariant at each its
step, permitting use of abstract representation of tensor 
quantities.

\subsection{Timelike Congruences}

The notion of reference frames is considered in the scope of classical
physics, namely in the assumption that the observation and measurement
procedures do not disturb the spacetime geometry. This means that the
reference bodies are test ones and the only limitation on their motion
is due to the relativistic causality principle: since a reference body
represents an idealisation of sets of measuring devices and local
observers, its' points worldlines should be timelike. Therefore we
shall assume that the motion of a reference body is described by a
congruence of timelike worldlines.

It is also obviously possible to an observer to move together with a
frame of reference, locally or in a spacetime region (globally). When
an object moves together with a reference frame, it is geometrically
identified with the latter, so that the worldlines of its mass points
form a congruence. Such a congruence presents a complete
characterisation of the reference frame, and with any reference frame
a conceptual object (thus a test one) of the above type can be
associated, which is called the body of reference. Since it models a
set of observers and their measuring devices, but no photons, the
reference frame congruence should be timelike.

We conclude that in the spacetime region where such a reference frame
is realised, we consider a congruence of integral curves of a unit
timelike vector field which is naturally interpreted as a field of
4-velocities of local observers, or equivalently of the worldlines of
particles forming a reference body. Every local observer represents
such a test particle.

It is important to note that the congruence concept is essential
because for the sake of regularity of the mathematical description of
the frame, these lines have not to mutually intersect, and they must
cover completely the spacetime region under consideration, so that at
every world point one has to find one and only one line passing
through it. The simplest way to describe a reference frame is to
identify the congruence of local observers with a congruence of the
time coordinate lines, which should be timelike [32].

Thus all the tensor quantities, as well as all the differential
operators, defined on the spacetime manifold, can be projected onto the
physical time direction of a frame of reference and its local
three-space with the help of appropriate projectors.

Let us assume that $u$ is a future pointing unit timelike vector field
($u^a u_a =-1$) representing the 4-velocity field of a family of test
observers filling the spacetime or some open submanifold of it.

The observer-orhogonal decomposition or [1+3] decomposition of the
tangent space, and, in turn of the algebra of spacetime tensor fields,
is accomplished by the temporal projection operator $v(u)$ along $u$
and the spatial projection operator $h(u)$ onto the orthogonal local
rest space, which may be identified with mixed second rank tensors
acting by contraction
\begin {equation}
{\delta^a}_b={v(u)^a}_{b} + {h(u)^a}_{b}
\end {equation}
where
\begin {equation}
{v(u)^a}_{b} = -u^a u_b
\end {equation}
\begin {equation}
 {h(u)^a}_{b}={\delta^a}_b + u^a u_b 
\end {equation}

These satisfy the usual orthogonal projection relations 
\begin {equation}
v(u)^2=v(u), \quad h(u)^2=h(u)
\end {equation}
and
\begin {equation}
v(u)h(u)=h(u)v(u)=0
\end {equation}

If $S$ is a general tensor, then the ``measurement'' of $S$ by the
observer congruence is the family of spatial tensor fields which
result from the spatial projection of all possible contractions of $S$
by any number of factors of $u$. For example, if $S$ is a
(1,1)-tensor, then its measurement
\begin {equation}
{S^a}_b \rightarrow (u^d u_c {S^c}_d, {h(u)^a}_{c} u^d {S^c}_d, {h(u)^d}_{a} u_c {S^c}_d, {h(u)^a}_{c}{h(u)^d}_{b} {S^c}_d)
\end {equation}
results in a scalar field, a spatial vector field, a spatial 1-form
and a spatial (1,1)-tensor field. It is exactly this family of fields
which occur in the orthogonal decomposition of $S$ with respect to the
observer congruence
\begin {eqnarray}
{S^a}_b&=&[{v(u)^a}_{c}+ {h(u)^a}_{c}][{v(u)^d}_{b}+ {h(u)^d}_{b}]  {S^c}_d \nonumber \\
       &=&[u^d u_c {S^c}_d]u^a u_b + \ldots + {h(u)^a}_{c}{h(u)^d}_{b} {S^c}_d
\end{eqnarray}

The reference frame generated by $u^a$ is described in the theory of
timelike congruences by the introduction of its kinematical quantities
$$ \sigma_{ab},\quad \omega_{ab},\quad \theta,\quad \dot{u_a}$$
which are defined by the irreducible decomposition or ``measurement''
of the covariant derivative $u_{a;b}$ with respect to $u^a$ [22,23]
\begin {equation}
u_{a;b}=\sigma_{ab} +  \omega_{ab} +\frac{1}{3} \theta h_{ab} -\dot{u_a}u_b
\end {equation}
where 
$\dot{u_a}$ denotes the acceleration vector of reference frame;
$\theta$ the volume rate of expansion scalar; $\sigma_{ab}$ is the rate
of shear tensor, with magnitude $$\sigma^2=\frac{1}{2}\sigma_{ab}
\sigma^{ab} $$ and   $\omega_{ab}$ is the vorticity tensor. It is
convenient to define a vorticity vector $$ \omega^a= \frac{1}{2}
\eta^{abcd}   \omega_{bc} u_d$$ denoting the angular velocity vector of
frame of reference where $\eta^{abcd}$ is the totally skew-symmetric
permutation tensor.

We note that that an overdot over a kernel letter means derivation
with respect to $u_a$, hence $\dot{u_a}=u_{a;b}u^b$.

These concepts were borrowed by the references frames theory from
hydrodynamics where they play an important role. They all are also
important in the theory of null congruences, often used in the
classification of gravitational fields by principal null directions
(the Petrov types) and also generation of exact Einstein-Maxwell
solutions [33].

Finally, we note that a partial splitting of spacetime based only on
a timelike congruence (splitting off time alone) is referred as the
congruence or ``time plus space'' or [1+3] decomposition, whereas a
spacelike slicing of spacetime (splitting off space alone) is referred
as the hypersurface splitting or ``space plus time'' or [3+1]
decomposition [34,35,36]. The two formalisms coincide in the case of
the observer-orthogonal decomposition of the tangent space of the
spacetime manifold.

If we had chosen the vector field generating a spacetime symmetry to
be timelike, then the covariant formalism provided by the theory of
timelike congruences would be enough to develop our method. Instead,
we have chosen the more complicated case in which we have a spacelike
symmetry vector. In this case a more sophisticated covariant formalism
is needed and we are naturally led to use the concept of a double congruence.

\subsection{Double Congruences}

For our purposes we consider a double congruence which involves two
vector fields: a timelike vector field $u^a$ representing the
4-velocity of a family of test observers and a spacelike vector field
$\xi^a$, which corresponds to a physical observable vector field, for
example electric field or magnetic field. We demonstrate the previous
point by considering the electromagnetic field strength tensor
$F_{ab}=F_{[ab]}$.

Then $$E^a={F^a}_b u^b \quad and \quad H^a=\frac{1}{2} \eta^{abcd} u_b
F_{cd}$$
Hence $$E^a u_a=H^a u_a=0$$

and both the electric and magnetic field as measured by the observers
$u^a$ have spacelike character.

We set $\xi^a=\xi \eta^a$, where  $\eta^a$ is a unit spacelike vector
$\eta^a \eta_a=1$ normal to the 4-velocity vector $u^a$.

To observe the given spacelike curves generated by $\eta$ in the
vicinity of the spacetime point $P$, we introduce, at that point, an
observer moving  with a 4-velocity $u^a$. We further suppose that the
spacelike curve $C$ is orthogonal to $u^a$ at the point $P$; then $$u^a \eta_a=0$$
It is important to emphasize that given a spacelike vector there is
not a unique, orthogonal, timelike unit vector associated with it. We
may add to the vector $u^a$ any vector $t^a$ such that $$
w^a=u^a+t^a$$
where $t^a$ satisfies the conditions
$$ t_a u^a=0 \quad and \quad t_at^a+2t_au^a=0$$
This freedom in our choice of an observer is essential in the
covariant character of the theory.

In what follows we shall restrict attention to the observer moving
with a 4-velocity $u^a$. Furthermore for the purpose of observation
this observer erects a screen orthogonal to the spacelike curve $C$ at
$P$. That is, the congruence of curves passes perpendicularly through
the screen.

Except at the given point $P$, the motions of the observers employed
along the curve $C$ have still to be specified. We require that the
4-velocities $u^a$ of the observers used along $C$ are related by:
\begin {equation}
{p^a}_c {u_c}^{\ast}={p^a}_c \dot{\eta^c}
\end {equation}

\begin {equation}
(u_a u^a)^{\ast}=0, \quad (u_a \eta^a)^{\ast}=0
\end {equation}

where an asterisk denotes derivation with respect to
$\eta^a$, hence
 $$ {u^c}^{\ast}=u_{a;b}\eta^b$$

The above ensures that $u^a$ is always a unit vector orthogonal to $\eta^a$ along $C$. Equations (14) and (15) are equivalent to the
single condition
\begin {equation}
{u_a}^{\ast}= \dot{\eta^a}+(\dot{\eta^b} u^b)u^a-({\eta_b}^{\ast} u^b) \eta^a
\end {equation}
which we call the Greenberg's transport law for $\eta^a$ [25].

The decomposition with respect to the double comgruence $(u,\eta)$ or
[1+1+2] decomposition of the tangent space, and, in turn of the
algebra of spacetime tensor fields, is accomplished by the temporal
projection operator $v(u)$ along $u$, the spatial projection operator
$s(\eta)$ along $\eta$, and the screen projection operator $p(u,\eta)$
which projects normally to both $(u,\eta)$ onto an orthogonal two
dimensional space, called the screen space.

All the above projection operators may be identified with mixed second
rank tensors acting by contraction.
\begin{equation}
{\delta^a}_b={v(u)^a}_b+{s(\eta)^a}_b+{p(u,\eta)^a}_b
\end {equation} 
\begin{equation}
{v(u)^a}_b=-u^a u_b
\end {equation} 
\begin{equation}
{s(u)^a}_b=\eta^a \eta_b
\end {equation}
\begin{equation} 
{p(u,\eta)^a}_b={\delta^a}_b+u^a u_b- \eta^a \eta_b={h^a}_b-\eta^a \eta_b
\end {equation}

The covariant derivative $\eta_{a;b}$ can be decomposed with respect
to the double congruence $(u,\eta)$ as follows
\begin{equation} 
\eta_{a;b}=A_{ab}+{\eta_a}^* \eta_b+u_a[\eta^tu_{t;b}+(\eta^t \dot u_t)u_b-(\eta^t{u_t}^*)\eta_b]
\end {equation} 
where
\begin{equation} 
A_{ab}={p^c}_a {p^d}_b \eta_{c;d}
\end {equation} 
We decompose $A_{ab}$ further into its irreducible parts with respect
to the orthogonal group:
\begin{equation} 
A_{ab}={\mathcal T}_{ab}+{\mathcal R}_{ab}+\frac{1}{2}{\mathcal E} p_{ab}(u,\eta)
\end {equation} 

where ${\mathcal T}_{ab}={\mathcal T}_{ba}$,${{\mathcal T}^a}_a=0 $ is
the traceless part of $A_{ab}$, and ${\mathcal R}_{ab}$ is the rotation
of  $A_{ab}$. We have the relations 
\begin{equation} 
{\mathcal T}_{ab}={p^c}_a {p^d}_b \eta_{(c;d)}-\frac{1}{2}{\mathcal E} p_{ab}
\end {equation} 
\begin{equation} 
{\mathcal R}_{ab}={p^c}_a {p^d}_b \eta_{[c;d]}
\end {equation} 
\begin{equation} 
{\mathcal E}=p^{cd} \eta_{c;d}={\eta^a}_{;a} + {\dot \eta}^a u_a
\end {equation}

The tensors ${\mathcal R}_{ab}$,${\mathcal T}_{ab}$ and the scalar
${\mathcal E}$ are defined as the kinematical quantities of the
spacelike congruence and have the following physical significance:
${\mathcal R}_{ab}$ represents the screen rotation,${\mathcal T}_{ab}$
the screen shear and ${\mathcal E}$ the screen expansion.

It is easy to show that in (16) the $u^a$ term in parentheses can be
written in a very useful form as follows:
$$ -N_b+2\omega_{tb}\eta^t +{p^t}_b {\dot \eta}_t$$
where  
\begin{equation} 
N_b={p^a}_b({\dot \eta}_a-{u_a}^{\ast})
\end {equation}

On using (22), equation (16) takes the form
\begin{equation} 
\eta_{a;b}=A_{ab}+{\eta_a}^* \eta_b+{p^c}_b {\dot \eta}_c u^a +
(2\omega_{tb}\eta^t -N_b) u_a
\end {equation}

The vector $N^a$, which is called Greenberg's vector, is of fundamental importance in the theory of double
congruences. Geometrically the condition $N^a=0$ means that the
congruences $u^a$ and $\eta^a$ are two surface forming. Kinematically,
it means that the field $\eta^a$ is ``frozen in'' along the observers
$u^a$.

We show in this work that the role of Greenberg's vector is more
general and establishes a connection between the field equations and
the symmetries at kinematical level.

Using (16) we can also prove the following useful identities that the
Lie derivatives of the projection tensors $p_{ab}$ and $h_{ab}$ obey
and which we
are going to use later
\begin {equation}
\frac{1}{\xi}L_{\xi}p_{ab}=2[{\mathcal T}_{ab}+\frac{1}{2}{\mathcal E}p_{ab}]-2u_{(a}N_{b)}
\end {equation}
\begin {equation}
\frac{1}{\xi}L_{\xi}h_{ab}=2[{\mathcal
T}_{ab}+\frac{1}{2}{\mathcal E}p_{ab}]-2u_{(a} N_{b)}+2(\log {\xi})_{(,a}
\eta_{b)}+2\eta_{(a}^{\ast} \eta_{b)}
\end {equation}

\section{Geometrisation of Spacetime Symmetries}

The types of symmetries we are going to deal with, in what follows are
those which arise from the existence of a Lie algebra of vector fields
on the spacetime manifold which are invariant vector fields of certain
geometrical objects on this manifold.

In Riemannian geometry the building block is the metric tensor
$g_{ab}$ in the sense that all the important geometrical objects of
this geometry are expressed in terms of $g_{ab}$.

Following the standard literature [2,16,27], we define the generic form of a
symmetry to be 
\begin{equation}
L_{\xi}g_{ab}=2 \Psi g_{ab} + H_{ab}
\end {equation}

where $\Psi$ is a scalar field and $H_{ab}$ is a symmetric traceless
tensor field. Both of the fields satisfy a unique set of conditions
specific to each particular symmetry and lead to a unique
decomposssition of $L_{\xi}g_{ab}$. As special examples we mention
that the Killing symmetries are characterized by $$\Psi=H_{ab}=0$$
the Homothetic symmetries by $$ \Psi=\phi=constant \quad and \quad
H_{ab}=0$$
the Conformal symmetries by $$\Psi=\omega(x^a) \quad and \quad
H_{ab}=0$$
whereas the Affine symmetries by $\Psi$, $H_{ab}$ such that
$$\Psi_{;c}=0 \quad and \quad H_{ab;c}=0$$

The generic form of a spacetime symmetry permits us to treat all of
them in a unifying manner and is essential to our approach.

The geometrisation of spacetime symmetries is managed if we describe it
as necessary and sufficient conditions on the geometry of the integral
lines of the vector field which generates the symmetry. For our
purposes we are going to study the geometrisation of a general
spacetime symmetry generated by a spacelike vector field $ \xi^a= \xi
\eta^a$ using the framework of the double congruence $(u, \eta)$
developed in section 2.

In the above framework the geometrisation of a a spacetime symmetry is
established through the following theorem:

{\bf Theorem}:The vector field  $ \xi^a= \xi \eta^a$  is a solution of
$$L_{\xi}g_{ab}=2 \psi g_{ab} + H_{ab}$$ if and only if

\begin{equation}
{\mathcal T}_{ab}=\frac{1}{2 \xi}({p^c}_a {p^d}_b -\frac{1}{2}
p^{cd}p_{ab})H_{cd}
\end{equation}
\begin{equation}
{\dot \eta}^a u_a=\frac{1}{\xi}(- \Psi) +\frac{1}{2 \xi} H_{11}
\end{equation}
\begin{equation}
 {\eta^a}^{\ast}=-u^a[- \dot {\log \xi}
+\frac{1}{\xi} H_{21}]+p^{ab}[- (\log \xi)_{,b} + \frac{1}{\xi} H_{b2}]
\end{equation}
\begin{equation}
 \xi^{\ast}= \psi + \frac{1}{2} H_{22}
\end{equation}
\begin{equation}
{\mathcal E}=\frac{2 \psi}{\xi} + \frac{1}{2 \xi}p^{ab} H_{ab}
\end{equation}
\begin{equation}
N_a=-2 \omega_{ab} \eta^b +\frac{1}{\xi} {p^b}_a H_{b1}
\end{equation}
where we use the notational convention $$Z \ldots _a \ldots u^a=Z \ldots _1 \ldots \quad and \quad Z \ldots _a \ldots \eta^a=Z \ldots _2 \ldots$$
for every tensor field $Z$.

{\bf Proof}

The equation
$$L_{\xi}g_{ab}=2 \Psi g_{ab} + H_{ab}$$
can be written equivalently in the form
\begin{equation}
\xi (\eta_{a;b} + \eta_{b;a}) +\xi_{,a} \eta_b +\xi_{,b} \eta_a=2\Psi g_{ab}+H_{ab}
\end{equation}
We decompose the above equation with respect to the double congruence
$(u,\eta)$. This can be done by contracting (26) with $u^a u^b$, $u^a
\eta^b$, $u^a {p^b}_c$, $\eta^a \eta^b$, $\eta^a {p^b}_c$ and ${p^a}_c
{p^b}_d$ respectively.

We obtain in turn:
\begin{equation}
u^a u^b:\quad {\dot \eta}^a u_a=\frac{1}{\xi}(- \Psi) +\frac{1}{2 \xi} H_{11}
\end{equation}
\begin{equation}
u^a
\eta^b:\quad  {\eta^a}^{\ast}u_a =- \dot {\log \xi}
+\frac{1}{\xi} H_{21}
\end{equation}
\begin{equation}
u^a {p^b}_c:\quad \xi {p^a}_c ({\dot \eta}_a + \eta_{b;a} u^b)={p^a}_c H_{a1}
\end{equation}
\begin{equation}
\eta^a \eta^b:\quad \xi^{\ast}=\Psi +\frac{1}{2} H_{22}
\end{equation}
\begin{equation}
\eta^a {p^b}_c:\quad {p^b}_c \eta_b^\ast={p^b}_c[-(\log \xi)_{,b} +\frac{1}{\xi} H_{b2}]
\end{equation}
\begin{equation}
{p^a}_c{p^b}_d:\quad {\mathcal T}_{cd}+\frac{1}{2} {\mathcal E}
p_{cd}=\frac{1}{2 \xi} {p^a}_c {p^b}_d H_{ab}+\frac{1}{\xi}\Psi p_{cd}
\end{equation}

In equation (36) the first term of the lhs can be written in the form
\begin {eqnarray}
\xi {p^a}_c ({\dot \eta}_a+\eta_{b;c}u^b)&=& \nonumber \\
\xi {p^a}_c({\dot \eta}_a-{u_a}^{\ast}+u_{a;b} \eta^b-u_{b;a} \eta^b)&=& \nonumber \\
\xi N_c+ \xi {p^a}_c (u_{a;b}-u_{b;a}) \eta^b&=& \nonumber \\
\xi N_c + 2\xi {p^a}_c \omega_{ab} \eta ^b
\end{eqnarray}

Substituting (40) in (36) and using $${p^b}_a \omega_{bc}
\eta^c=\omega_{ac} \eta^c$$ we obtain
\begin{equation}
N_a=-2 \omega_{ab} \eta^b + \frac{1}{\xi} {p^b}_a H_{b1}
\end{equation}

Equtions (35) and (38) give us the components of ${\eta^a}^*$ along
$u^a$ and on the screen space of $\eta^a$,$u^a$. It is possible to
combine them into a single equation
\begin{equation}
 {\eta^a}^{\ast}=-u^a[- \dot {\log \xi}
+\frac{1}{\xi} H_{21}]+p^{ab}[- (\log \xi)_{,b} + \frac{1}{\xi} H_{b2}]
\end{equation}

Next we consider the trace and the traceless part of (39) and we obtain
correspondingly
\begin{equation}
Trace:\quad{\mathcal E}=\frac{2 \Psi}{\xi} + \frac{1}{2 \xi}p^{ab} H_{ab}
\end{equation}
\begin{equation}
Traceless \quad part:\quad {\mathcal T}_{ab}=\frac{1}{2 \xi}({p^c}_a {p^d}_b -\frac{1}{2}
p^{cd}p_{ab})H_{cd}
\end{equation}

The converse of the theorem is proved as follows:

We consider the tensor
\begin{equation}
\xi_{(a;b)}-\Psi g_{ab} -\frac{1}{2}H_{ab}= \xi \eta_{(a;b)}+\xi_{(,a}
\eta_{b)} - \Psi g_{ab} -\frac{1}{2} H_{ab}
\end{equation}
Contracting this with $u^a u^b$, $u^a
\eta^b$, $u^a {p^b}_c$, $\eta^a \eta^b$, $\eta^a {p^b}_c$ and ${p^a}_c
{p^b}_d$ respectively and applying relations (27)-(32), we prove that
this tensor vanishes. The above completes the proof of the theorem.

The above theorem is of major importance to our approach because if we
use it the symmetry is expressed in a form that it is possible to be
incorporated in Einstein's field equations in a direct and inherent
way.

Besides the theorem can be used in other ways depending on the
information supplied. For example if information is available on the
vector field $\xi^a$ which generates the symmetry, one can investigate
what type of symmetry the given vector field can generate. Conversely
if information is available on the scalar and tensor fields
$\psi$,$H_{ab}$ one can use the theorem to obtain information on the
vector field $\xi^a$ which generates the symmetry.

\section{The symmetries-incorporated Einstein's equations}

The Einstein field equations with a non-zero cosmological constant can
be written as
\begin{equation}
R_{ab}=T_{ab}+(\Lambda-\frac{T}{2})g_{ab}
\end{equation}

where $\Lambda$ is the cosmological constant and $T$ is the trace of
the energy-momentum tensor $T_{ab}$.

We wish to incorporate a general spacetime symmetry into Einstein's
field equations. This is desirable because one can effectively
eliminate the symmetry from the field equations and find solutions
that will by construction comply with the symmetry.

The effects of the symmetries at the dynamical level are obtained by
the Lie derivation of the Einstein field equations 
\begin{equation} 
L_{\xi}R_{ab}=L_{\xi}[T_{ab}+(\Lambda-\frac{T}{2})g_{ab}]
\end{equation}

Eventually we obtain the field equations as Lie derivatives along the
symmetry vector. We assume that the symmetry vector $\xi^a$ is
spacelike orthogonal to the 4-velocity $u^a$ of the observers.

First of all we express the energy-momentum tensor $T_{ab}$ in terms
of its constituent dynamical variables by irreducibly decomposing it
with respect to the double congruence $(u, \eta)$. Next we compute the
rhs of (47) in terms of the Lie derivatives of the dynamical
variables. The next step is to compute the $L_{\xi}R_{ab}$ directly
from the generic form of the symmetry in terms of
$\xi_{a;b}$. Employing the $(u, \eta)$ double congruence framework we
use the expression of $\xi_{a;b}$ in terms of the kinematical
quantities, namely we use the geometrisation of the symmetry applying
the theorem proved in section 3. Thus we finally manage to write
the field equations in a way that they fully incorporate the character
of a spacetime symmetry.

\subsection{$[1+1+2]$ Irreducible decomposition of $T_{ab}$}

It is well known from the literature that the irreducible
decomposition of the energy momentum tensor $T_{ab}$ with respect to
the four velocity $u^a$ of observers, defines the dynamical
variables of spacetime [22,23] 
\begin{equation}
T_{ab}= \mu u_a u_b+ph_{ab}+2q_{(a}u_{b)}+\pi_{ab}
\end{equation}

where $\mu$ and $p$ denote the total energy density and the isotropic
pressure, $q_a$ is the heat flux vector, and $\pi_{ab}$ is the
traceless anisotropic stress tensor. These quantities include
contributions by all sources, for example by an electromagnetic
field. In particular  $q_a$ represents processes such as heat
conduction and diffusion as well as the electromagnetic flux.

We can further decompose irreducibly the
quantities  $q_a$ and  $\pi_{ab}$ using the double congruence
framework, since -except the timelike congruence of observers- there
is also defined a spacelike congruence on spacetime, representing
covariantly a physical observable field as well as the character of
the symmetry vector. 
\begin{equation}
q^a=v \eta^a+Q^a
\end{equation}
\begin{equation}
\pi_{ab}=\gamma(\eta_a \eta_b-\frac{1}{2}p_{ab})+2P_{(a} \eta_{b)} +D_{ab}
\end{equation}

where $$v=q^a \eta_a,\quad Q^a={p^a}_b q_b, \quad
\gamma=\pi_{ab}\eta^a \eta^b$$, $$P_a={p^a}_b {\pi^b}_c \eta^c, \quad
D_{ab}=({p^c}_a {p^d}_b-\frac{1}{2}p_{ab}p^{cd}) \pi_{cd}$$

The tensors $Q^a$, $P^a$, $D_{ab}$ are on the screen space of $(u,
\eta)$ and $D_{ab}$ is traceless.

The dynamical variables are constrained to obey the conservation
equations 
\begin{equation}
{T^{ab}}_{;b}=0
\end{equation}
which result from the identity ${G^{ab}}_{;b}=0$, where $G_{ab}$ denotes
the Einstein tensor. The above 
equations can be irrecucibly decomposed in the double congruence $(u,
\eta)$ framework into the following system of equations, which consist
the conservation laws that the dynamical variables of spacetime have
to satisfy.
\begin{equation}
\dot \mu+(\mu +p) \theta +{q^a}_{;a}+q^a \dot u^a + \pi^{ab} \sigma_{ab}=0
\end{equation}
\begin{eqnarray}
(\mu +p) (\dot u_a \eta^a)+(p_{;c}+\dot q_c+{\pi^b}_{c;b}) \eta^c
+\theta (q^a \eta_a) \nonumber \\
+q^b {u_b}^{\ast} -2({p^c}_b q^b) \omega_{bt} \eta^t =0
\end {eqnarray} 
\begin{eqnarray}
(\mu +p) \dot u_a p_{ac} +({p_{;}}^a +{\dot q}^a+ {\pi^{ba}}_{;b}) p_{ac}
\nonumber \\
+\theta q^a p_{ac} +q^b {u^a}_{;b} p_{ac} =0
\end {eqnarray}

\subsection{$[1+1+2]$ irreducible decomposition of the rhs of
Einstein's field equations}

\subsubsection{Computation of $L_\xi R_{ab}$ using the field equations}

We are going to compute the Lie derivative of the Ricci tensor using
the field equations. The field equations for general matter (46) using (48) can be
written in the form
\begin{equation}
R_{ab}=(\mu +p) u_a u_b +\frac{1}{2} (\mu -p+2 \Lambda) g_{ab}
+2q_{(a} u_{b)} +\pi_{ab}
\end{equation}

We study the case in which $\xi^a=\xi  \eta^a$ corresponding to $\xi^a$
spacelike and orthogonal to the 4-velocity of the observers. We
decompose the above equation with respect to the double congruence
$(u,\eta)$. This is achieved by contracting the Lie derivative of the
Ricci tensor with $u^a u^b$, $u^a
\eta^b$, $u^a {p^b}_c$, $\eta^a \eta^b$, $\eta^a {p^b}_c$ and ${p^a}_c
{p^d}_b$ respectively. If we denote by $\frac{1}{\xi} L_\xi
R_{ab}[uu]$, $\frac{1}{\xi} L_\xi R_{ab}[u \eta]$, $\frac{1}{\xi} L_\xi
R_{ab}[u p]$, $\frac{1}{\xi} L_\xi R_{ab}[\eta \eta]$, $\frac{1}{\xi}
L_\xi R_{ab}[\eta p]$, $\frac{1}{\xi} L_\xi R_{ab}[p p]$ the
independent projections correspondingly we obtain:   
\begin{eqnarray}
\frac{1}{\xi} L_\xi R_{ab}[uu]&=&[\frac{1}{2} (\mu +3p)^* +(\mu
+3p-2\Lambda)(\dot u^c \eta_c) \cr 
&-& 2(q^c \eta_c) \dot {(\log \xi)} -2(q^c N_c) +2(q^c
\eta_c)({u^d}^{\ast} \eta_d)]u_a u_b
\end{eqnarray}
\begin{eqnarray}
\frac{1}{\xi} L_\xi R_{ab}[u \eta]&=&- 2[-\frac{1}{2}(\mu -p
+2\Lambda)({u^c}^{\ast} \eta_c)  
-{(q^c \eta_c)}^* \pi_{cd}
\eta^c({u^d}^{\ast} - \dot \eta^d) \cr
&+& (q^c \eta_c)(\dot \eta^d u_d)
+ \frac{1}{2}(\mu -p +2\Lambda +2 \pi_{cd} \eta^c \eta^d) \dot {(\log
\xi)} \cr
&-&(q^c \eta_c) {(\log \xi)}^*]u_{(a} \eta_{b)}
\end{eqnarray} 
\begin{eqnarray}
\frac{1}{\xi} L_\xi R_{ab}[u p]&=&- 2[\frac{1}{2}(\mu -p +2\Lambda)N_d + (\mu +3p-2\Lambda)
\omega_{cd} \eta^c  \cr
&-&\pi_{ce} {p^c}_d ({u^e}^{\ast} - \dot \eta^e)-q_c A_{cd}-
{q^c}^{\ast} p_{cd} \cr
&+& {p^c}_d q_c (\dot \eta^e u_e)-(q^c \eta_c) {p^e}_d {(\log \xi)}_{;e}+
\eta^e {p^c}_d \pi_{ec} \dot {(\log \xi)}]u_{(a}{p^d}_{b)} 
\end{eqnarray} 
\begin{equation}
\frac{1}{\xi} L_\xi R_{ab}[\eta \eta]= [\frac{1}{2}(\mu -p +2 \pi_{cd} \eta^c \eta^d)^* + (\mu -p
+2\Lambda +2 \pi_{cd} \eta^c \eta^d) (\log \xi)^*] \eta_a \eta_b
\end{equation}
\begin{eqnarray}
\frac{1}{\xi} L_\xi R_{ab}[\eta p]&=& 2 [\frac{1}{2}(\mu -p +2\Lambda)p_{cd} \eta^{c*} +{p^c}_d
(\pi_{ce} \eta^e)^* \cr
&+&{\pi^c}_e \eta^e A_{cd} +2(q_t \eta^t)
\omega_{dc} \eta^c 
+ \eta^c {p^e}_d \pi_{ce} (\log \xi)^*  \cr
&+&\frac{1}{2}(\mu -p +2
\Lambda +2 \pi_{ce} \eta^c \eta^e){p^f}_d {(\log \xi)}_{,f}]\eta_{(a}
{p^d}_{b)} 
\end{eqnarray}
\begin{eqnarray}
\frac{1}{\xi} L_\xi R_{ab}[p p]&=&
 [(\mu -p +2\Lambda) ({\mathcal T}_{cd}+\frac{1}{2} {\mathcal E} 
p_{cd}) +\frac{1}{2}{(\mu-p)}^* p_{cd} \cr
&+&{\pi^e}_f(A_{ec}{p_d}^f
+A_{ed}{p_c}^f) 
+ {\pi^{ef}}^* p_{ec} p_{fd} \cr 
&+&4q_c{p^e}_{(c} \omega_{d)f} \eta^f +2
\eta^e \pi_{ef} {p_c}^{(f}{p^{t)}}_d {(\log \xi)}_{,t}]{p^c}_{(a}{p^d}_{b)} 
\end{eqnarray}

The above projections of the Lie derivative of the Ricci tensor have been
obtained  taking into account the irreducible
decomposition of the symmetric $(0,2)$ tensor $T_{ab}$ with respect to
$u^a$. However, since  $\xi^a$ is spacelike we have
further irreducible decompositions of the tensor fields. Concretely if we use
equations (49) and (50) we express the Lie derivative of the Ricci
tensor in terms of the irreducible parts of the dynamical
variables. Then the projections obtain the following irreducible form:
\begin{eqnarray}
\frac{1}{\xi} L_\xi R_{ab}[uu]&=&[\frac{1}{2} (\mu +3p)^* +(\mu
+3p-2\Lambda)(\dot u^c \eta_c) \cr
&-&2Q^c N_c -2v[\dot {(\log
\xi)}+\eta^{c*}u_c]]u_a u_b 
\end{eqnarray}
\begin{eqnarray}
\frac{1}{\xi} L_\xi R_{ab}[u \eta]&=&- 2[\frac{1}{2}(\mu -p +2\Lambda+2 \gamma)[\dot {(\log
\xi)}+\eta^{c*}u_c] \cr
&+& P^c N_c-v^*-v[ {(\log
\xi)}^{\ast}-\dot \eta^{c}u_c]]u_{(a} \eta_{b)}
\end{eqnarray}
\begin{eqnarray}
\frac{1}{\xi} L_\xi R_{ab}[u p]&=&
-2[\frac{1}{2}(\mu -p +2\Lambda- \gamma)N_c +(\mu
+3p-2\Lambda) \omega_{dc} \eta^d \cr
&+& P_c[\dot {(\log
\xi)}+\eta^{c*}u_c] 
+D_{dc}N^d - v{p^d}_c[{\eta_d}^* +{\log \xi}_{,d}] \cr
&+& Q_c(\dot \eta^d u_d)-Q_d
{A^d}_c -{p^d}_c Q_d^*]{p^c}_{(a}u_{b)} 
\end{eqnarray}
\begin{eqnarray}
\frac{1}{\xi} L_\xi R_{ab}[\eta \eta]=
 \frac{1}{2} [(\mu -p +2 \gamma)^* +(\mu -p +2\Lambda+2 \gamma)
{(\log \xi)}^* ] \eta_a \eta_b 
\end{eqnarray}
\begin{eqnarray}
\frac{1}{\xi} L_\xi R_{ab}[p \eta]&=&
 2[p_{dc} P^* +\frac{1}{2}(\mu -p +2\Lambda+2 \gamma){p^d}_c
[{\eta_d}^* +{(\log \xi)}_{,d}] \cr
&+& P^d [A_{dc}+{(\log \xi)}^* p_{cd}]-2v \omega_{dc} \eta^d]
{p^c}_{(a} \eta_{b)}
\end{eqnarray}
\begin{eqnarray}
\frac{1}{\xi} L_\xi R_{ab}[p p]&=&
 \frac{1}{2}[{(\mu -p - \gamma)}^* + (\mu -p
+2\Lambda-\gamma) {\mathcal E}]p_{ab} \cr
&+& {p^c}_a {p^d}_b {D_{cd}}^* +2A_{c(b}{D^c}_{a)}+(\mu -p
+2\Lambda-\gamma) {\mathcal T}_{ab} \cr
&+&4Q_{(a} \omega_{b)c} \eta^c
+2P_{(a} {p^c}_{b)} [{\eta_c}^* + {(\log \xi)}_{,c}]
\end{eqnarray}

\subsubsection{Incorporation of symmetries}

Using the theorem proved in section 3  we reexpress the Lie
derivative of the Ricci tensor in terms of the quantities $\Psi$,
$H_{ab}$ which characterise the generic form of a spacetime symmetry. 
Thus we finally obtain the independent projections of the Lie derivative of the Ricci tensor along the
symmetry generating vector in terms of the fields characterising the symmetry $\Psi$,$H_{ab}$ and the irreducible
dynamical variables in the following form: 

\begin{eqnarray}
\frac{1}{\xi} L_\xi R_{ab}[uu]&=&[\frac{1}{2} (\mu +3p)^* +(\mu
+3p-2\Lambda) (\frac{\Psi}{\xi}
-\frac{1}{2 \xi} H_{11}) \cr
&-&\frac{2v}{\xi} H_{21}-2Q^c N^c]u_a u_b
\end{eqnarray}
\begin{eqnarray}
\frac{1}{\xi} L_\xi R_{ab}[u \eta]&=&
-2[\frac{1}{2 \xi} (\mu -p+2\Lambda+2\gamma)H_{21} \cr
&+&P_c
N^c-v^*-v\frac{1}{2\xi}(4\Psi +H_{22}-H_{11})]u_{(a} \eta_{b)}
\end{eqnarray}
\begin{eqnarray} 
\frac{1}{\xi} L_\xi R_{ab}[\eta \eta]&=&
[\frac{1}{2}{[( \mu-p+2\Lambda)+2\gamma]}^* \cr
&+&[(
\mu-p+2\Lambda)+2\gamma]{(\log \xi)}^*] \eta_a \eta_b 
\end{eqnarray}
\begin{eqnarray} 
\frac{1}{\xi} L_\xi R_{ab}[pu]&=&
-2[\frac{1}{2}[(\mu-p+2\Lambda)+2\gamma]N_c+(\mu+3p-2\Lambda)\omega_{tc}\eta^t-{Q_t}^*
{p^t}_c \cr
&-&\frac{v}{\xi} H_{t2}{p^t}_c-Q^t H_{tc} 
+ D_{tc}N^t +\frac{1}{\xi} H_{21}P_c-Q^t {\mathcal R}_{tc} \cr
&-&Q_c(\frac{2\Psi}{\xi}
+\frac{1}{4\xi}p^{ef}
H_{ef}-\frac{1}{2\xi}H_{11})]{p^c}_{(a}u_{b)} 
\end{eqnarray}
\begin{eqnarray} 
\frac{1}{\xi} L_\xi R_{ab}[p \eta]&=&
 2[{p^d}_c {P_d}^* +P_c(\frac{2\Psi}{\xi}+\frac{1}{4\xi}p^{ef}
H_{ef}+\frac{1}{2\xi}H_{22})+P_d {H^d}_c +P_d {{\mathcal R}^d}_c
\nonumber \\
&+& \frac{1}{2\xi}[( \mu-p+2\Lambda)+2\gamma]{p^d}_c H_{d2}-2v \eta^d
\omega_{dc}]{p^c}_{(a} \eta_{b)} 
\end{eqnarray}
\begin{eqnarray} 
\frac{1}{\xi} L_\xi R_{ab}[pp]&=&
\frac{1}{2}{(\mu-p-\gamma)}^*p_{ab}+[(
\mu-p+2\Lambda)+2\gamma][H_{ab} \cr
&+&\frac{1}{4\xi}p^{cd} H_{cd} p_{ab}
+\frac{\Psi}{\xi} p_{ab}]  
+ 2D_{c(a} {{\mathcal R}^c}_{b)} +{\mathcal E} D_{ab} +2D_{c(a}
{H_{b)}}^c \cr
&+&{p^c}_{(a}{p^d}_{b)} {D_{cd}}^*+ \frac{2}{\xi} P_{(a}
{p^c}_{b)} H_{c2}+4Q_{(a} \omega_{b)t} \eta^t
\end{eqnarray}

\subsection{$[1+1+2]$ irreducible decomposition of the lhs of
Einstein's field equations}

We consider the generic form of a spacetime symmetry
$$ L_\xi g_{ab}=2\Psi g_{ab}+H_{ab}$$
where $${H^a}_a=0 \quad and \quad H_{ab}=H_{ba}$$

The Lie derivative of the connection coefficients in terms of the Lie
derivative of the metric tensor is expressed as follows [4]:  
\begin{equation}
L_\xi {\Gamma^a}_{bc}=\frac{1}{2} g^{ad}[(L_\xi g_{ab})_{;c}+(L_\xi g_{ac})_{;b}-
(L_\xi g_{bc})_{;d}]
\end{equation}

Combining equations (26) and (74) we obtain directly:
\begin{equation}
L_\xi {\Gamma^a}_{bc}=\frac{1}{2} g^{ad}[2\Psi_{;c} g_{bd} +2\Psi_{;b} g_{dc}- 
2\Psi_{;d} g_{b} +H_{db;c}+H_{dc;b}+H_{bc;d}]
\end{equation}

Equation (75) is equivalent to the following:
\begin{equation}
L_\xi {\Gamma^a}_{bc}=\Psi_{;c}{\delta^a}_b+\Psi_{;b}{\delta^a}_c-{\Psi_{;}}^a{g^b}_c+\frac{1}{2}[H_{db;c}+H_{dc;b}+H_{bc;d}]
\end{equation}

Furthermore it is easy to obtain
\begin{equation}
L_\xi {\Gamma^a}_{ab}=4\Psi_{;b}
\end{equation}

In order to calculate the Lie derivative of the Ricci tensor we apply
the relation
\begin{equation}
L_\xi R_{ab}={(L_\xi{\Gamma^s}_{ab})}{;s}-{(L_\xi{\Gamma^s}_{as})}{;b}
\end{equation}

Hence we obtain:
\begin{eqnarray}
L_\xi R_{ab}&=&{[\Psi_{;a}{\delta^s}_b+\Psi_{;b}{\delta^s}_a
-{\Psi_{;}}^s
g_{ab}+\frac{1}{2}({H^s}_{a;b}+{H^s}_{b;a}-{H_{ab;}}^s)]}_{;s}-4 \Psi_{;a}
\nonumber \\
&=& (\psi_{;ab}+\Psi_{b;a}-{{\psi_;}^s}_s g_{ab}-4\Psi_{;ab})+\frac{1}{2}({{H^s}_a}_{;bs}+{{H^s}_b}_{;as}+{{{H_{ab}}_{;}}^s}_s)
\end {eqnarray}

Equivalently we obtain:
\begin{equation}
L_\xi R_{ab}=-2\Psi_{;ab}-\Box \psi g_{ab}+\frac{1}{2}({{H^s}_a}_{;bs}+{{H^s}_b}_{;as}+{{{H_{ab}}_{;}}^s}_s)
\end {equation}

or
\begin{equation}
L_\xi R_{ab}=-2\Psi_{;ab}-\Box \psi g_{ab}+\frac{1}{2} \Lambda_{ab}
\end {equation}
where
\begin{equation}
\Lambda_{ab}={{H^s}_a}_{;bs}+{{H^s}_b}_{;as}+{{{H_{ab}}_{;}}^s}_s
\end {equation}

Next we decompose $\Psi_{;ab}$ with respect to the double congruence
$(u,\eta)$ framework into its irreducible parts
\begin{eqnarray}
\Psi_{;ab}&=&\lambda _{\Psi} u_a u_b-2k_\Psi \eta_{(a} u_{b)}-2s_{\Psi(a} u_{b)}
\nonumber \\
&+& \gamma_\Psi \eta_a \eta_b +2p_{\Psi(a}\eta_{b)}+D_{\Psi
ab}+\frac{1}{2} a_\Psi p_{ab} 
\end {eqnarray}

where
$$\lambda _{\Psi}=\Psi_{;ab} u^a u^b,\quad k_\Psi =\Psi_{;ab}\eta^{(a}
u^{b)},\quad s_{\Psi a}={p_a}^b \Psi_{;bc} u^c$$
$$ \gamma_\Psi=\Psi_{;ab}\eta^a \eta^b,\quad 2p_{\Psi a}= {p_a}^b
\Psi_{;bc} \eta^c $$ $$ D_{\Psi ab}=({p^c}_a
{p^d}_b-\frac{1}{2}p_{ab}p^{cd}) \Psi_{;cd},\quad a_\Psi=\Psi_{;ab} p^{ab}$$
Moreover we have
$$g_{ab} \Psi_{;ab}=-\lambda _{\Psi}+\gamma_\Psi+ a_\Psi$$

Thus the sum $-2\Psi_{;ab}-\Box \psi g_{ab}$ is decomposed irreducibly
as
\begin{eqnarray}
-2\Psi_{;ab}-\Box \psi g_{ab}&=& (-3\lambda _{\Psi}+\gamma_\Psi+
a_\Psi)u_a u_b +4k_\Psi \eta_{(a} u_{b)} \nonumber \\
&+& 4s_{\Psi (a} u_{b)} +(\lambda _{\Psi}-3\gamma_\Psi-a_\Psi) \eta_a
\eta_b -4p_{\Psi(a}\eta_{b)} \nonumber \\
&-&2D_{\Psi ab} +(\lambda _{\Psi}-\gamma_\Psi-2a_\Psi)p_{ab} 
\end{eqnarray}

Similarly the tensor $\Lambda_{ab}$ defined by equation (82) is
decomposed irreducibly as follows:
\begin{eqnarray}
\frac{1}{2} \Lambda_{ab}&=&\lambda_\Lambda u_a u_b-2k_\Lambda \eta_{(a} u_{b)}-2
s_{\Lambda (a} u_{b)} \nonumber \\
&+& \gamma_\Lambda \eta_a \eta_b +2p_{\Lambda(a} \eta_{b)} +D_{\Lambda
ab} +\frac{1}{2} a_\Lambda p_{ab}
\end{eqnarray}

Thus the lhs of Einstein's equations is decomposed irreducibly as
follows:
\begin{equation}
\frac{1}{\xi} L_\xi R_{ab}[uu]=\frac{1}{\xi} (\lambda_\Lambda -3\lambda
_{\Psi}+\gamma_\Psi+a_\Psi)u_a u_b 
\end{equation}
\begin{equation}
\frac{1}{\xi} L_\xi R_{ab}[u \eta]=\frac{1}{\xi} 
(-2(k_\Lambda-2k_\Psi))
\eta_{(a}u_{b)}
\end{equation}
\begin{equation}
\frac{1}{\xi} L_\xi R_{ab}[up]=\frac{1}{\xi} 
[-2(s_{\Lambda c}-2s_{\Psi c})] {p^c}_{(a}
u_{b)}
\end{equation}
\begin{equation}
\frac{1}{\xi} L_\xi R_{ab}[\eta \eta]=\frac{1}{\xi} 
[(\gamma_\Lambda-3\gamma_\Psi)+(\lambda_\Psi-a_\Psi)] \eta_a
\eta_b 
\end{equation}
\begin{equation}
\frac{1}{\xi} L_\xi R_{ab}[\eta p]=\frac{1}{\xi} 
 [2(p_{\Lambda c}-2p_{\Psi c})]{p^c}_{(a} \eta_{b)}
\end{equation}
\begin{equation}
\frac{1}{\xi} L_\xi R_{ab}[p p]= \frac{1}{\xi} 
[D_{\Lambda ab}
-2D_{\Psi ab}+ (\lambda_\Psi-\gamma_\Psi-2a_{\Psi}+\frac{1}{2}
a_\Lambda) p_{ab}]
\end{equation}

\subsection{The symmetries-incorporated field equations for the
irreducible dynamical variables}

Our purpose is to construct the field equations for general matter 
that  the dynamical variables $\mu,p,\gamma,v,Q^a,P^a,D_{ab}$ of
spacetime satisfy, in such a way that the
information of  any particular spacetime symmetry imposed, is directly
incorporated in the form of the equations. As a first step we equate
the results we have obtained previously for the rhs and the lhs of
Einstein's equations:  
\begin{eqnarray}
&[\frac{1}{2} (\mu +3p)^* +(\mu
+3p-2\Lambda)(\frac{\Psi}{\xi}-\frac{1}{2 \xi} H_{11}) 
-\frac{2v}{\xi} H_{21}-2Q^c N^c]=  \nonumber \\
&\frac{1}{\xi} (\lambda_\Lambda -3\lambda
_{\Psi}+\gamma_\Psi+a_\Psi)
\end{eqnarray}

\begin{eqnarray}
&[\frac{1}{2}{[(
\mu-p+2\Lambda)+2\gamma]}^*+[(\mu-p+2\Lambda)+2\gamma](\frac{\Psi}{\xi}+\frac{1}{2\xi}H_{22})]=
\nonumber \\
&\frac{1}{\xi} ((\gamma_\Lambda-3\gamma_\Psi)+(\lambda_\Psi-a_\Psi)) 
\end{eqnarray}

\begin{eqnarray}
&[\frac{1}{2 \xi} (\mu -p+2\Lambda+2\gamma]H_{21} +P_c
N^c-v^*-v\frac{1}{2\xi}(4\Psi +H_{22}-H_{11})=  \nonumber \\
&\frac{1}{\xi}  (k_\Lambda-2k_\Psi)
\end{eqnarray}

\begin{eqnarray}
&[\frac{1}{2}[(\mu-p+2\Lambda)-\gamma]N_c+(\mu+3p-2\Lambda)\omega_{tc}\eta^t-{Q_t}^*
{p^t}_c -\frac{v}{\xi} H_{t2}{p^t}_c-Q^t H_{tc} \nonumber \\
&+ D_{tc}N^t +\frac{1}{\xi} H_{21}P_c-Q^t {\mathcal R}_{tc}
-Q_c(\frac{2\Psi}{\xi}+\frac{1}{4\xi}p^{ef}
H_{ef}-\frac{1}{2\xi}H_{11})] \nonumber \\
&=\frac{1}{\xi} (s_{\Lambda c}-2s_{\Psi c})  
\end{eqnarray}

\begin{eqnarray}
&[{p^d}_c {P_d}^* +P_c(\frac{2\Psi}{\xi}+\frac{1}{4\xi}p^{ef}
H_{ef}+\frac{1}{2\xi}H_{22})+P_d {H^d}_c +P_d {{\mathcal R}^d}_c
\nonumber \\
&+ \frac{1}{2\xi}[( \mu-p+2\Lambda)+2\gamma]{p^d}_c H_{d2}-2v \eta^d
\omega_{dc}] \nonumber \\
&=\frac{1}{\xi}(p_{\Lambda c}-2p_{\Psi c})  
\end{eqnarray}

\begin{eqnarray}
&{(\mu-p-\gamma)}^*+(\mu-p+2\Lambda-\gamma)(\frac{1}{2\xi}
p^{ef}H_{ef}+\frac{2\Psi}{\xi}) \nonumber \\
&2D^{cd}H_{cd} +\frac{2}{\xi}P^c H_{c2}+4Q^c \omega_{ct} \eta^t
\nonumber \\
&=2(\lambda_\Psi-\gamma_\Psi-2a_{\Psi}+\frac{1}{2}
a_\Lambda) 
\end{eqnarray}

\begin{eqnarray}
&{p^c}_a{p^d}_b{D_{cd}}^* +{\mathcal E} D_{ab}
+(\mu-p+2\Lambda-\gamma)H_{ab}+2D_{c(a} {{\mathcal
R}^c}_{b)} \nonumber \\
&+2({p^c}_{(a}{p^d}_{b)}-\frac{1}{2} p_{ab}p^{cd})(D_{ec}
{H^e}_d+2Q_{(c} \omega_{d)t} \eta^t) \nonumber \\
&=D_{\Lambda ab}-2D_{\Psi ab}
\end{eqnarray}

The above system of equations can give us the irreducible equations
that each of the dynamical variables of spacetime obey, if we manage
to disentagle it through appropriate algebraic manipulations.

Equations (92),(93) and (97) can be written equivalently as follows correspondingly:
\begin{eqnarray}
&\mu^*+(\mu+\Lambda)(\frac{2\Psi}{\xi}-\frac{1}{\xi}H_{11})+3(p^*+(p-\Lambda))(\frac{2\Psi}{\xi}-\frac{1}{\xi}H_{11})
\nonumber \\
&=
\frac{2}{\xi}(\lambda_\Lambda-3\lambda_\Psi+\gamma_\Psi+a_\Psi)+\frac{4v}{\xi}H_{21}+4Q_c
N^c
\end{eqnarray}

\begin{eqnarray}
&\mu^*+(\mu+\Lambda)(\frac{2\Psi}{\xi}+\frac{1}{\xi}H_{22})-[p^*+(p-\Lambda)(\frac{2\Psi}{\xi}+\frac{1}{\xi}H_{22})]
\nonumber \\
&
+2[\gamma^*+\gamma(\frac{2\Psi}{\xi}-\frac{1}{\xi}H_{22})]=\frac{2}{\xi}[\lambda_\Psi
-a_\Psi+(\gamma_\Lambda-3\gamma_\Psi)]
\end{eqnarray}

\begin{eqnarray}
&\mu^*+(\mu+\Lambda)(\frac{2\Psi}{\xi}+\frac{1}{2\xi}p^{ef}H_{ef})-[p^*+(p-\Lambda)(\frac{2\Psi}{\xi}+\frac{1}{2\xi}p^{ef}H_{ef})]
\nonumber \\
&-[\gamma^*+\gamma(\frac{2\Psi}{\xi}+\frac{1}{2\xi}p^{ef}H_{ef})]=-4Q^c
\omega_{ct} \eta^t \nonumber \\
&+\frac{2}{\xi}(\lambda_\Psi-\gamma_\Psi-2a_\Psi+\frac{1}{2}a_\Lambda)-2D^{cd}H_{cd}-\frac{2}{\xi}P^c
H_{c2}
\end{eqnarray}

Moreover we note that
$$g^{ab}H_{ab}=0 \quad  or \quad p^{ab}H_{ab}=H_{11}-H_{22}$$

Hence equation (101) can be written in the form
\begin{eqnarray}
&\mu^*+(\mu+\Lambda)[\frac{2\Psi}{\xi}+\frac{1}{2\xi}(H_{11}-H_{22})]-[p^*+(p-\Lambda)(\frac{2\Psi}{\xi}+\frac{1}{2\xi}(H_{11}-H_{22}))]
\nonumber \\
&-[\gamma^*+\gamma(\frac{2\Psi}{\xi}+\frac{1}{2\xi}(H_{11}-H_{22}))]=-4Q^c
\omega_{ct} \eta^t \nonumber \\
&+\frac{2}{\xi}(\lambda_\Psi-\gamma_\Psi-2a_\Psi+\frac{1}{2}a_\Lambda)-2D^{cd}H_{cd}-\frac{2}{\xi}P^c
H_{c2}
\end{eqnarray}

Next if (102) is multiplied by a factor of two and added to (100) gives
\begin{eqnarray}
&\mu^*+(\mu+\Lambda)(\frac{2\Psi}{\xi}+\frac{1}{3\xi}H_{11})-(p^*+(p-\Lambda)(\frac{2\Psi}{\xi}+\frac{1}{3\xi}H_{11}))
\nonumber \\
&+\gamma(\frac{1}{\xi}H_{22}-\frac{1}{3\xi}H_{11})=\frac{1}{3\xi}(6\lambda_\Psi-10a_\Psi+2\gamma_\Lambda-10\gamma_\Psi+2a_\Lambda)
\nonumber \\
& -\frac{8}{3}Q^c \omega_{ct} \eta^t-\frac{4}{3}
D^{cd}H_{cd}-\frac{4}{3\xi}P^c H_{c2}
\end{eqnarray}

Furthermore we multiply (103) by a factor of three and we add to (99)
\begin{eqnarray}
&\mu^*+(\mu+\Lambda)\frac{2\Psi}{\xi}-(p-\Lambda)
\frac{1}{\xi}H_{11}+\frac{1}{4\xi} \gamma(3H_{22}-H_{11}) \nonumber \\
&=\frac{1}{2\xi}[\lambda_\Lambda+\gamma_\Lambda+a_\Lambda-4\gamma_\Psi-4a_\Psi]+\frac{v}{\xi}H_{21}+Q_c
N^c \nonumber \\
&-2Q^c \omega_{ct} \eta^t -D^{cd} H_{cd}-\frac{1}{\xi}P^cH_{c2}
\end{eqnarray}

Next we consider the difference of (99) and (103)
\begin{eqnarray}
&p^*+(p-\Lambda)(\frac{2\Psi}{\xi}-\frac{2}{3\xi} H_{11})-\frac{1}{4\xi}
\gamma (H_{22}-\frac{1}{3}H_{11})-\frac{1}{3\xi}(\mu+\Lambda)H_{11}
\nonumber \\
&=\frac{1}{6\xi}(3\lambda_\Lambda-\gamma_\Lambda-a_\Lambda+8a_\Psi+8\gamma_\Psi-12
\lambda_\Psi)+Q_c N^c \nonumber \\
&+\frac{2}{3} Q^c \omega_{ct} \eta^t +\frac{1}{3}
D^{cd}H_{cd}+\frac{1}{3\xi}P^c H_{c2} +\frac{v}{\xi}H_{21}
\end{eqnarray}

Finally we subtract (102) from (100) and we obtain: 
\begin{eqnarray}
&[\gamma^*+\gamma(\frac{2\Psi}{\xi}+\frac{1}{6\xi}H_{11}+\frac{1}{2\xi}H_{22})]+(\mu+\Lambda)
\frac{1}{2\xi}(-\frac{1}{\xi}H_{11}+H_{22}) \nonumber \\
&+(p-\Lambda)\frac{1}{2\xi}(\frac{1}{\xi}H_{11}-H_{22})=\frac{2}{3\xi}(\gamma_\Lambda-\frac{1}{2}a_\Lambda
-2\gamma_\Psi+a_\Psi) \nonumber \\
&+\frac{2}{3} D^{cd} H_{cd} +\frac{2}{3} \frac{1}{\xi} P^c H_{c2} +\frac{4}{\xi}Q^c \omega_{ct} \eta^t
\end{eqnarray}

Thus the final irreducible form of the symmetries-incorporated
Einstein field equations for the dynamical variables resulting from
the decomposition of the energy-momentum tensor for general matter is
described from the following system of equations:

{\bf Symmetries-incorporated field equations for $\mu, p, \gamma$}

\begin{eqnarray}
&\mu^*+(\mu+\Lambda)\frac{2\Psi}{\xi}-(p-\Lambda)
\frac{1}{\xi}H_{11}+\frac{1}{4\xi} \gamma(3H_{22}-H_{11}) \nonumber \\
&=\frac{1}{2\xi}[\lambda_\Lambda+\gamma_\Lambda+a_\Lambda-4\gamma_\Psi-4a_\Psi]+\frac{v}{\xi}H_{21}+Q_c
N^c \nonumber \\
&-2Q^c \omega_{ct} \eta^t -D^{cd} H_{cd}-\frac{1}{\xi}P^cH_{c2}
\end{eqnarray}

\begin{eqnarray}
&p^*+(p-\Lambda)(\frac{2\Psi}{\xi}-\frac{2}{3\xi} H_{11})-\frac{1}{4\xi}
\gamma (H_{22}-\frac{1}{3}H_{11})-\frac{1}{3\xi}(\mu+\Lambda)H_{11}
\nonumber \\
&=\frac{1}{6\xi}(3\lambda_\Lambda-\gamma_\Lambda-a_\Lambda+8a_\Psi+8\gamma_\Psi-12
\lambda_\Psi)+Q_c N^c \nonumber \\
&+\frac{2}{3} Q^c \omega_{ct} \eta^t +\frac{1}{3}
D^{cd}H_{cd}+\frac{1}{3\xi}P^c H_{c2} +\frac{v}{\xi}H_{21}
\end{eqnarray}

\begin{eqnarray}
&[\gamma^*+\gamma(\frac{2\Psi}{\xi}+\frac{1}{6\xi}H_{11}+\frac{1}{2\xi}H_{22}]+(\mu+\Lambda)
\frac{1}{2\xi}(-\frac{1}{\xi}H_{11}+H_{22}) \nonumber \\
&+(p-\Lambda)\frac{1}{2\xi}(\frac{1}{\xi}H_{11}-H_{22})=\frac{2}{3\xi}(\gamma_\Lambda-\frac{1}{2}a_\Lambda
-2\gamma_\Psi+a_\Psi) \nonumber \\
&+\frac{2}{3} \frac{1}{\xi} P^c H_{c2} +\frac{4}{\xi}Q^c \omega_{ct}
\eta^t +\frac{2}{3} D^{cd} H_{cd}
\end{eqnarray}

{\bf Symmetries-incorporated field equations for $v, Q^a, P^a, D_{ab}$}

\begin{eqnarray}
&v^*+v\frac{1}{2\xi}(4\Psi +H_{22}-H_{11})
 \nonumber \\
&=[-\frac{1}{2 \xi} (\mu -p+2\Lambda+2\gamma)]H_{21} +P_c
N^c
-\frac{1}{\xi}  (k_\Lambda-2k_\Psi)
\end{eqnarray}

\begin{eqnarray}
&{Q_t}^*
{p^t}_c +Q^t H_{tc}
+Q^t {\mathcal R}_{tc}
+Q_c(\frac{2\Psi}{\xi}-\frac{1}{4\xi}(H_{11}+H_{22})] \nonumber \\
&=-\frac{1}{\xi} (s_{\Lambda c}-2s_{\Psi c})-
[+\frac{1}{2}[(\mu-p+2\Lambda)-\gamma]N_c+(\mu+3p-2\Lambda)\omega_{tc}\eta^t
\nonumber \\
&-\frac{v}{\xi} H_{t2}{p^t}_c+ D_{tc}N^t +\frac{1}{\xi} H_{21}P_c
\end{eqnarray}

\begin{eqnarray}
&{p^d}_c {P_d}^* +P_c[\frac{2\Psi}{\xi}+\frac{1}{4\xi}(H_{11}+H_{22})]+P_d {H^d}_c +P_d {{\mathcal R}^d}_c
\nonumber \\
&=- \frac{1}{2\xi}[( \mu-p+2\Lambda)+2\gamma]{p^d}_c H_{d2}+2v \eta^c
\omega_{dc} \nonumber \\
&+\frac{1}{\xi}(p_{\Lambda c}-2p_{\Psi c})  
\end{eqnarray}

\begin{eqnarray}
&{p^c}_a{p^d}_b{D_{cd}}^* +{\mathcal E} D_{ab}
+2D_{c(a} {{\mathcal
R}^c}_{b)} \nonumber \\
&+2({p^c}_{(a}{p^d}_{b)}-\frac{1}{2} p_{ab}p^{cd})D_{ec}
{H^e}_d \nonumber \\
&=D_{\Lambda ab}-2D_{\Psi ab}-(\mu-p+2\Lambda-\gamma)H_{ab}-4Q_{(c} \omega_{d)t} \eta^t({p^c}_{(a}{p^d}_{b)}-\frac{1}{2} p_{ab}p^{cd})
\end{eqnarray}

The system of equations (107)-(113) provide the desirable irreducible
decomposition of Einstein's equations for general matter when an
arbitrary symmetry has been introduced in spacetime, such that the
information regarding the symmetry is explicitly contained in the
field equations.

This system of equations provides the key formal results of the paper
and generalises all the previous attempts to attack the problem, in an
elegant and unifying manner. Moreover it greatly enlarges the scope of
previous works since it can be applied to all types of symmetries as
well as to all types of matter. Thus it consists the unified
geometrical framework that we have to take into account when
discussing Einstein's equations in spacetimes with general
symmetries. Aside from elegance and covariance properties it provides the direct link among seperate approaches discussing
particular symmetries and matter fields.

It is evident that the exact physical significance of the various
parameters involved in the equations is provided by the special type
of symmetry and matter that somebody considers applying the
formalism. It is also expectable that the system of the
symmetries-incorporated Einstein equations will reduce to familiar
forms when applied to well-studied specific matter fields as well as
symmetries. Even in this case there are some new insights to be gained
from seeing these old calculations in the new general setting.

In the following section we demonstrate how these formal results are
used in the case of a spacelike Conformal Killing vector symmetry,
discussing the perfect fluid case. We expect to explore the
applications of the symmetries-incorporated Einstein equations in the
situation of Affine and Curvature collineations for various types of
matter in later papers.

\section{Application : The case of a spacelike conformal Killing
vector symmetry}

\subsection{Kinematical Level}

A spacelike conformal Killing vector  $\xi^a=\xi \eta^a (\eta^a
u_a=0)$ satisfies the equation
\begin{equation}
L_\xi g_{ab}=2\Psi g_{ab}
\end{equation}
We note that in this case $H_{ab}=0$.

From the theorem proved in section 3 the geometrisation of this
particular symmetry is described by the following set of conditions:
\begin{equation}
{\mathcal T}_{ab}=0
\end{equation}
\begin{equation}
{\eta_a}^*+{(\log \xi)}_{,a}=\frac{1}{2} {\mathcal E} \eta_a
\end{equation}
\begin{equation}
\dot{\eta_a} u^a=-\frac{1}{2} {\mathcal E}
\end{equation}
\begin{equation}
{p^b}_a(\dot {\eta_b}+u^t \eta_{t;b})=0
\end{equation}

The conformal factor $\Psi$ reads
\begin{equation}
\Psi=\frac{1}{2} \xi {\mathcal E}={\xi}^*
\end{equation}

Due to the condition $u^a \eta_a=0$ equation (118) can be written in the form
\begin{equation}
N_a=-2 \omega_{ab} \eta^b
\end{equation}

Moreover the decomposition of the covariant derivative of the
spacelike vector field $\eta_a$ is written as
\begin{equation}
\eta_{a;b}=A_{ab}+{\eta_a}^* \eta_b- \dot{\eta_a} u_b +{p^c}_b
\dot{\eta_c} u_a
\end{equation}
from which immediately obtain
\begin{equation}
L_\xi \eta^a=-\psi \eta^a
\end{equation}

Moreover using (120) we easily find
\begin{equation}
L_\xi u^a=-\Psi u^a-\xi N^a
\end{equation}

From (24) and (25) we can also compute
\begin{equation}
L_\xi p_{ab}=2\Psi p_{ab}-2\xi u_{(a} N_{b)}
\end{equation}

\begin{equation}
L_\xi h_{ab}=2\Psi h_{ab}-2\xi u_{(a} N_{b)}
\end{equation}

If $N^a=0$ then $(u, \eta)$ span a two dimensional surface in
spacetime. The screen space is the orthogonal complement to this
surface. The screen space admits $\xi^a$ as a spacelike conformal
Killing vector with conformal factor $\Psi$ and its metric is defined
by $p_{ab}(u,\eta)$. The rest space of the observers is the three
dimensional space normal to $u^a$. The rest space admits admits
$\xi^a$ as a spacelike conformal Killing vector with conformal factor
$\Psi$ and its metric is defined by $h_{ab}(u)$.

Finally, for a conformal Killing vector we have from (80) the following
further equations
\begin{equation}
L_\xi R_{ab}=-2\Psi_{;ab}-g_{ab} \Box \Psi
\end{equation}
\begin{equation}
L_\xi R=-2\Psi R-6 \Box \Psi 
\end{equation}

\subsection{Dynamical Level}

We use the set of equations (107)-(113) in the special case we study and
the results from the kinematical level. It is straightforward to obtain the following field equations:

\begin{equation}
\mu^*+(\mu+\Lambda) {\mathcal E}=-\frac{2}{\xi}(a_\Psi+\gamma_\Psi)+2Q_aN^a
\end{equation}
\begin{equation}
p^*+(p-\Lambda){\mathcal E}=\frac{2}{3\xi}(2a_\Psi+2\gamma_\Psi-3\lambda_\Psi)+\frac{2}{3}Q_aN^a
\end{equation}
\begin{equation}
\gamma^*+\gamma{\mathcal E}=-\frac{4}{3\xi}(\gamma_\Psi-\frac{1}{2} a_\Psi)-\frac{2}{3}Q_aN^a
\end{equation}
\begin{equation}
v^*+v{\mathcal E}=\frac{2}{\xi}k_\Psi +P_a N^a
\end{equation}
\begin{equation}
{p^d}_c {Q^d}^* +{\mathcal E}Q_c=\frac{2}{\xi}s_{\Psi
c}+(\mu+p-\frac{1}{2}\gamma)N_c+D_{dc}-Q_d {{\mathcal R}^d}_c
\end{equation}
\begin{equation}
{p^d}_c {P_d}^* +{\mathcal E}P_c=-\frac{2}{\xi}P_{\Psi c}+v N_c
-P_d {{\mathcal R}^d}_c
\end{equation}
\begin{equation}
{p^c}_a {p^d}_b{D_{cd}}^*+{\mathcal E}D_{ab}=-\frac{2}{\xi}D_{\Psi
ab}-2{\mathcal R}_{c(a} {D^c}_{b)}+2({p^c}_a
{p^d}_b-\frac{1}{2}p^{cd} p_{ab}) Q_{(c} N_{d)}
\end{equation}

The dynamics is fully specified if in addition to the field equations
we consider the conservation equations (52)-(54), which if expressed in
terms of their ireducible parts read

\begin{eqnarray}
& \dot \mu+(\mu+p-\frac{\gamma}{2}) \theta +v^*+2v {\mathcal E}
+p^{ab} Q_{a;b} +Q^a {(\log \xi)}_{;a} \nonumber \\
&+2Q^a \dot{u_a}+\frac{3}{2}\gamma \dot{(\log \xi)} +2 \sigma_{ab}P^a \eta^b+\sigma_{ab}D^{ab}=0
\end{eqnarray}

\begin{eqnarray}
&(p+\gamma)^*+\frac{1}{2}(\mu+p+4\gamma){\mathcal E}+ \dot v \nonumber
\\
&+v[\theta+ \dot{(\log \xi)} ]+p^{ab}P_{a;b}+P_a(\dot {u^a}- {\eta^a}^*)=0
\end{eqnarray}

\begin{eqnarray}
&{p^c}_a[(\mu+p-\gamma/2) \dot{u_c} +p_{;c}+ \dot{Q_c}+\frac{4}{3}
\theta Q_c+\sigma_{cd}Q^d \nonumber \\
&+2v\sigma_{cd} \eta^d +{P_c}^* +{\mathcal R}_{cd} P^d +{\mathcal E}
P_c +{{D_c}^d}_{;d}-\frac{3}{2}\gamma{(\log \xi)}_{;c}-\frac{1}{2}\gamma_{;c}]=0
\end{eqnarray}

The system of equations (128)-(137) characterise completely the dynamics
induced by the Einstein equations when a spacelike Conformal Killing
vector symmetry exists in spacetime. Thus it is possible if we specify
a concrete type of matter field and after introducing appropriate
systems of coordinates to obtain solutions which by construction
comply with this symmetry.

Since all the dynamical information is contained in (128)-(137) all of the
propositions or no-go theorems which can arise in the above setting,
and gradually have appeared in the literature using various methods,
are contained implicitly in the system we have constructed. In order
to clarify this point, in what follows, we specialise our discussion
in the case of a perfect fluid spacetime.  

\subsection{The Perfect fluid case}

In the case of a perfect fluid the following relations hold:
\begin{equation}
\psi_{;ab}=\lambda_\Psi u_a u_b +\gamma_\Psi h_{ab}+\xi (\mu+p)N_{(a} u_{b)}
\end{equation}
\begin{equation}
\Box \Psi=3\gamma_\Psi-\lambda_\Psi
\end{equation}

\begin{eqnarray}
&L_\xi R_{ab}=3(\gamma_\Psi-\lambda_\Psi) u_a u_b \nonumber \\
&+(-5\gamma_\Psi+\lambda_\Psi)h_{ab}-2\xi(\mu+p)N_{(a}u_{b)}
\end{eqnarray}
\begin{equation}
L_\xi R=-2\Psi R-6(3\gamma_\Psi-\lambda_\Psi)
\end{equation}

The general field equations in the case of a perfect fluid when
a conformal Killing vector symmetry is introduced take the form:
\begin{equation}
\mu^*+(\mu+\Lambda) {\mathcal E}=-\frac{6}{\xi} \gamma_\Psi
\end{equation}
\begin{equation}
p^* +(p-\Lambda) {\mathcal E}=\frac{2}{\xi}(2\gamma_\Psi-\lambda_\Psi)
\end{equation}
\begin{equation}
(\mu+p)N_a=-\frac{2}{\xi}s_{\Psi a}
\end{equation}
\begin{equation}
0=k_\Psi=P_{\Psi a}=D_{\Psi ab}
\end{equation}
\begin{equation}
a_\Psi=2\gamma_\Psi
\end{equation}

The consevation equations become
\begin{equation}
p^*+\frac{1}{2}(\mu+p) {\mathcal E}=0
\end{equation}
\begin{equation}
(\mu-p+2\Lambda \Psi=4\gamma_\Psi+2\lambda_\Psi
\end{equation}
\begin{equation}
(\mu+p){p^b}_a \dot{u_b}+{p^c}_ap_{,c}=0
\end{equation}

Using the above sets of equations we can recover all the theorems that have
been proved in the literature in a large series of publications using
other methods, quite easily. Since the large majority of these
theorems are well known we are not going to state all of them, but we
will restrict ourselves in mentioning two examples of propositions of
this kind, so as to prove the increased flexibility of our formalism
in familiar situations, and at the same time, to gain new insights by
their embedding in a unified geometrical framework.

{\it Proposition: }Let $\xi^a$ be a proper homothetic spacelike
Killing vector orthogonal to the 4-velocity of observers of a perfect fluid
spacetime ($\Psi_{;a}=0$, $\Psi\neq 0$). Then $p-\Lambda=\mu+\Lambda$,
namely matter is stiff and the current $j^a:={\xi^{[a;b]}};b$
vanishes.

{\it Proof: } If $\xi^a$ is a conformal Killing vector we can easily
obtain that the vector field $\Psi_{;a}$ obeys the following identity:
\begin{eqnarray}
\Psi_{;a}&=&-\frac{1}{3}(\mu+p)(u_t \xi^t)u_a +
\frac{1}{6}(p-\mu-2\Lambda) \xi^a \cr
&-&\frac{2}{3}q_{(a}u_{b)} \xi^b- \frac{1}{3} \pi_{ab} \xi^b
-\frac{1}{3} j^a
\end{eqnarray}

The above identity results if we apply the Ricci identity and use the
expression of the Ricci tensor in terms of the dynamical variables
from the field equations.

Since we refer to a perfect fluid spacetime the above equation reads
\begin{equation}
(2\gamma_\Psi-\lambda_\Psi)\xi_a=\Psi(3\Psi_{;a}+j^a)
\end{equation}

Next we use (143) and we immediately see that the condition
$\Psi_{;a}$ implies that $p-\Lambda=\mu+\Lambda$, or else matter is stiff and the current $j^a:={\xi^{[a;b]}};b$
vanishes.

{\it Proposition: }Let $\xi^a$ be a spacelike conformal Killing vector
orthogonal to the 4-velocity of observers of a perfect fluid
spacetime. Then if $\Lambda=0$, $\mu+p=0$ and the dominant energy
condition holds, spacetime is a de Sitter spacetime with constant
positive curvature and $\xi^a$ reduces to a Killing vector.

{\it Proof: }From conservation equations we find that $\mu=const$ and
from our assumptions positive since $\mu+p=0$. Field equations (142)
and (143)
imply $\gamma_\Psi=-\lambda_\Psi$ and field equation (144) that $s_{\Psi
a}=0$.
From these results we have:
\begin{equation}
\Psi_{;ab}=\gamma_{\Psi}g_{ab}
\end{equation}
\begin{equation}
L_\xi R_{ab}=-\gamma_\Psi g_{ab}
\end{equation}
\begin{equation}
L_\xi R=-2 \Psi R -24\gamma-\Psi
\end{equation}
From Einstein's field equations we find in this case:
\begin{equation}
R_{ab}=\mu g_{ab}
\end{equation}
\begin{equation}
R=4\mu=const
\end{equation}
Thus spacetime is a de-Sitter spacetime. Moreover equation (155), (156) give
\begin{equation}
\mu \Psi=-3\gamma_\Psi
\end{equation}
Applying Ricci's identity to the vector $\Psi_{;a}$ we find
$R^{ab}\Psi_{;a}=0$ which upon replacing $R_{ab}$ from (155) we obtain
$\mu \Psi_{;a}=0$. But we have $\mu>0$ hence the above gives
$\Psi_{;a}=0$ or $\gamma_\Psi=0$. Thus from (157) we conclude that
$\Psi=0$, hence $\xi^a$ reduces to a Killing vector.

\section{Summary and Discussion}

The basic tool to simplify Einstein's field equations, in search for
exact solutions, has been the introduction of spacetime
symmetries. The latter form Lie algebras of vector fields on the
spacetime manifold which are invariant vector fields of certain
geometrical objects on  this manifold.

Related to the initial motivation, it would be  very desirable to
have in hand a general framework permitting the incorporation of
a symmetry of every possible kind in Einstein's equations for general
matter. There lacks in the literature a unified approach that will
apply to all the symmetries and also to general matter. We have
presented a general method to handle the field equations for general
matter when a spacetime symmetry is introduced in its work.

Our method has been expressed in a covariant formalism, using the
framework of a double congruence $(u,\eta)$, permitting the
introduction of arbitrary reference frames and eventually systems of
coordinates. The basic notion on which it is based is that of the
geometrisation of a general symmetry, namely the description of it as
an equivalent set of conditions on the kinematical quantities
characterising the congruence of the integral lines of the vector
field that generates the symmetry.

Using the above notion we finally manage to write the field equations
which the dynamical variables obey for any type of matter, in a way
that they inherit the symmetry of the generating vector.

The method has been applied in the case of a spacelike Conformal Killing
vector symmetry recovering completely the existing literature. Further
applications have been considered recovering results obtained by other
methods in the case of a perfect fluid spacetime.

Thus we finally generalise and extend the results of the current
literature in the form of the symmetries-incorporated Einstein's field
equations for general matter.

The usefulness of such a construction lays on the following facts:

Firstly the field equations for general matter obtain an irreducible
form with respect to the covariant congruence framework and  specify
completely the evolution of the spacetime dynamical variables 
inheriting, at the same time, the symmetry of the vector that generates it.

Secondly the system of the symmeties-incorporated Einstein equations
give us the opportunity to eliminate the symmetry from the field
equations trivially and find exact solutions that will by construction
comply with the symmetry.

Thirdly, the set of the symmetries-incorporated field equations can
be also used in another perspective equally significant. That is the
above set of equations can be considered as a system of integrability
conditions for the existence of a spacetime symmetry of a particular
kind, in spacetime, when a concrete form of the energy-momentum tensor
is specified. Conversely it is possible imposing a spacetime symmetry
to examine what types of matter can be present in spacetime. Related
to the above remark the existing literature, using other methods, give
us the information that some of the discussed symmetries are either
absent or tightly restricted if a specific form of the energy momentum
tensor is given. For example, affine and conformal vector fields
cannot exist in vaccum spacetimes 
and affines are also forbidden when we have a perfect fluid with $0
\leq p \neq \rho >0$ and are also severely restricted for
Einstein-Maxwell spacetimes [16,37]. Curvature symmetries are also heavily
restricted in a similar way [38]. Along these lines research is in
progress, using the formalism developed in this paper.

We finally wish to mention a  short remark, which shows another
fruitful direction  in the same framework.

The method presented in this paper can be also used to study the
symmetry inheritance of a kinematical or dynamical variable. This
concept has been defined in the literature [29,31,39] to mean that in the
presence of a spacetime symmetry a kinematical or dynamical variable $X$,
satisfies an equation of the form $$L_\xi X+k\psi X=0$$ where $k$ is a
constant depending on the tensorial character of $X$. It is an
appealing idea because it relates the symmetry with all the variables
kinematical and dynamical making full use of the field equations.

\vspace{1.5cm}
\noindent{\bf Aknowledgements}

\bigskip

It is my great pleasure to thank M. Tsamparlis for many stimulating
discussions, as well as G.S. Hall and C. Casares. I would also
especially  like to thank C. Isham and J.J. Halliwell for their
comments. This work has been partially supported by A.S. Onassis
public benefit foundation.
v

\end{document}